# Self-reconfigurable Multifunctional Memristive Nociceptor for Intelligent Robotics

Shengbo Wang,[1,2] Mingchao Fang,[1] Lekai Song,[3] Cong Li,[1] Jian Zhang,[1] Arokia Nathan,[4] Guohua Hu,[3] and Shuo Gao[1]

[1]*School of Instrumentation and Optoelectronic Engineering, Beihang University, Beijing, 100191, China*

[2]*Tsinghua-Berkeley Shenzhen Institute, Tsinghua University, Beijing, 100084, China*

[3]*Department of Electronic Engineering, The Chinese University of Hong Kong, Hong Kong, 999077, China*

[4]*Darwin College, Cambridge University, Cambridge, CB3 9EU, UK*

Artificial nociceptors, mimicking human-like stimuli perception, are of significance for intelligent robotics to work in hazardous and dynamic scenarios. One of the most essential characteristics of the human nociceptor is its self-adjustable attribute, which indicates that the threshold of determination of a potentially hazardous stimulus relies on environmental knowledge. This critical attribute has been currently omitted, but it is highly desired for artificial nociceptors. Inspired by these shortcomings, this article presents, for the first time, a Self-Directed Channel (SDC) memristor-based self-reconfigurable nociceptor, capable of perceiving hazardous pressure stimuli under different temperatures and demonstrates key features of tactile nociceptors, including 'threshold,' 'no-adaptation,' and 'sensitization.' The maximum amplification of hazardous external stimuli is 1000%, and its response characteristics dynamically adapt to current temperature conditions by automatically altering the generated modulation schemes for the memristor. The maximum difference ratio of the response of memristors at different temperatures is 500%, and this adaptability closely mimics the functions of biological tactile nociceptors, resulting in accurate danger perception in various conditions. Beyond temperature adaptation, this memristor-based nociceptor has the potential to integrate different sensory modalities by applying various sensors, thereby achieving human-like perception capabilities in real-world environments.

**1. Introduction**

After entering the era of artificial intelligence and autonomy, robotics are stepping into our lives in tremendous ways, including in industry, healthcare, and social service. A growing expectation of robotics is to conduct tasks traditionally reserved for human beings.[1–4] However, a significant challenge resulting in the mission being temporally 'out-of-reach' is that current intelligent robotics cannot perceive environmental information efficiently as humans do.[5,6] To this end, various electronic senses are developed and equipped into robotics, mimicking intricate human bionic sensing capabilities.[7,8] Nevertheless, the performances of robotics in many real-world scenarios are not satisfying yet, driving the motivation for further study on creating electronic copies of human perception mechanisms for robotics.[9,10]

Recent biomedical studies unveil that receptors are crucial in assisting biological perception systems to understand external information. These receptors, such as fast-adapting receptors, slow-adapting receptors, and nociceptors, allow for the perception and processing of external physical stimuli, culminating in nociception and sensory adaptation functionalities, which are significant for humans to tackle with structured and unstructured data.[11–13] Nociceptors, in particular, have triggered the special interest of robotic researchers due to their ability to spontaneously learn hazard signals, thereby enhancing the safety of robotics in environments with potential dangers.[14–17] This insight has given rise to fruitful and promising research outcomes. Notably, neuromorphic devices, similar to the basic computing units of biological systems, have been widely used to develop artificial nociceptors.[18–23]

In state-of-the-art results, functions such as 'threshold,' 'no-adaptation,' 'relaxation,' and 'sensitization' of nociceptors have been successfully obtained and employed.[11,24–26] For example, Dahiya et al. demonstrated a synaptic transistor that mimics an artificial nociceptor, equipping robotics with the function of pain reflex.[27] However, the response characteristics of these artificial nociceptors are fixed. In contrast, biological nociceptors have the ability to adjust their response characteristics based on current environmental conditions, resulting in perceptual capabilities with adapting and harmony functions.[28–31] This means that the fixed response characteristics of artificial nociceptors could make current techniques fail to respond to hazardous information accurately, particularly in dynamic environments. To address this issue, in this article, we introduce a self-reconfigurable memristor-based nociceptor for the first time. Utilizing the inherent similarity of memristors to biological synapses and modulation scheme design, this nociceptor exhibits a self-reconfigurable capability to detect dangerous pressure stimuli at various temperatures, and its ability to amplify dangerous stimuli is positively correlated with temperature, as shown in FIG 1. This behaviour closely mimics the functions of biological nociceptors, making a significant advancement in electronic analog of human perception mechanisms, and could further enhance the working performance of robotics in unstructured environments.

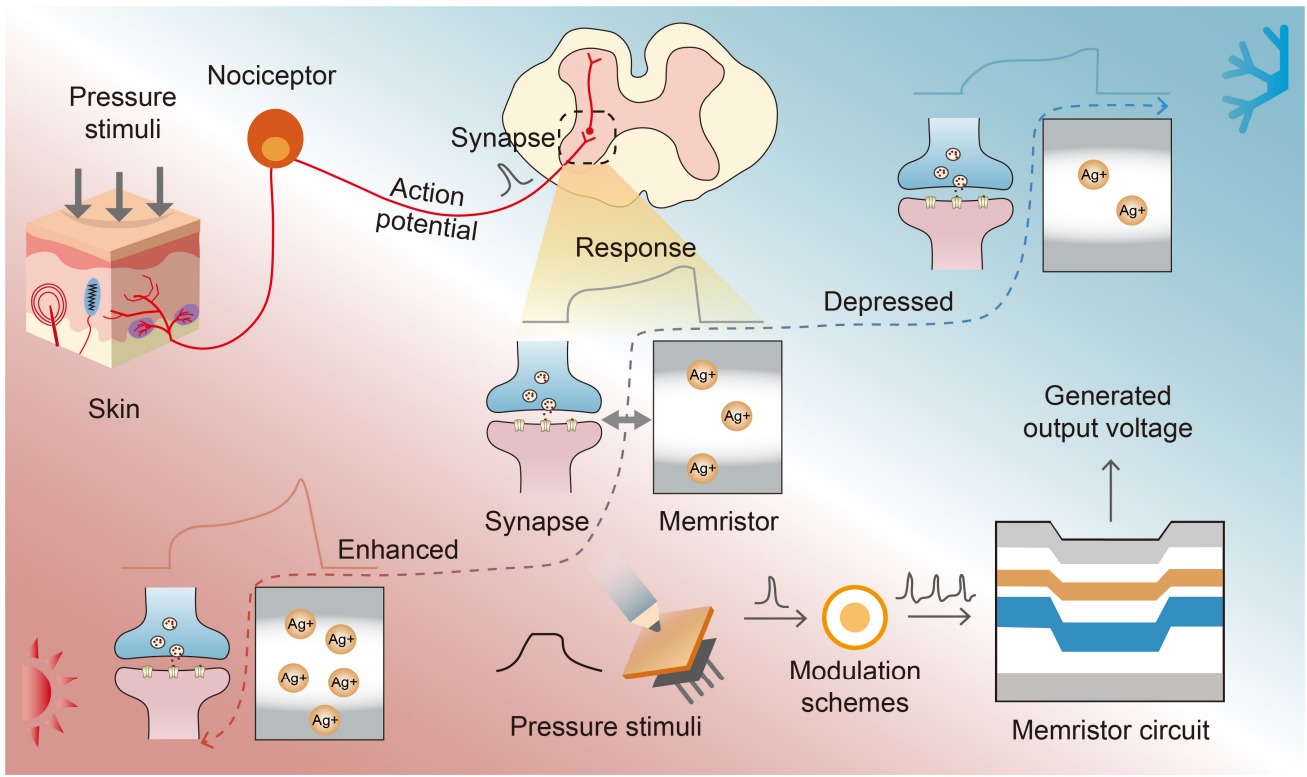

FIG.1. Schematic illustrations of a memristive nociceptor with adjustable response characteristics. The response of this memristive nociceptor to dangerous pressure stimuli is enhanced with rising temperatures and depressed with decreasing temperatures, respectively, closely mirroring that of biological tactile nociceptors.

## 2. Results and Discussion

### 2.1. Self-Directed Channel (SDC) Memristors

A self-directed channel (SDC) memristor (KNOWM Inc.) is selected for constructing this adjustable nociceptor. The SDC memristor follows a multi-layer structure as depicted in FIG 2a. In the initial operation, $Ge_2Se_3/Ag/Ge_2Se_3$ these three layers mix together to form the Ag source layer. Besides, Sn ions are generated from the SnSe layer and then incorporated into the active $Ge_2Se_3$ layer, leading to the creation of a pair of self-trapped electrons in the $Ge_2Se_3$ layer. As a result, Sn ions enable the energetically favourable reaction of Ag substituting for Ge on the Ge-Ge bonds; in other words, this reaction generates 'openings' for $Ag^+$ ions, which serve natural conductive channels within the active layer. During device operation, Ag can move onto or away from these agglomeration sites formed in conductive channels, as the resistance switching mechanism in SDC memristors (FIG 2a). Electrical measurements were performed using Tektronix Keithley parameter analyser and the hysteresis curve and pulse test is recorded as FIG 2b and FIG 2c. Such a wide resistance switching range, non-volatile characteristics and a certain threshold voltage have indicated that the SDC memristor is a potential candidate to construct artificial nociceptors.

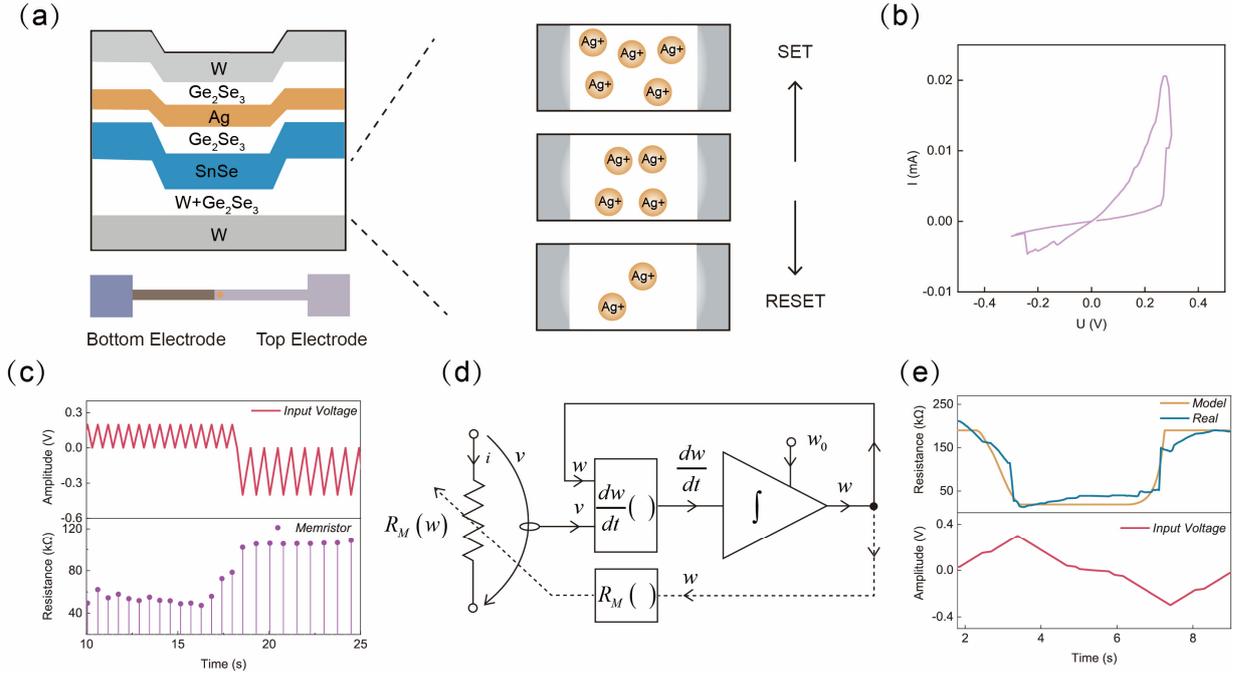

FIG.2. The self-directed channel (SDC) memristor and its electrical model. (a) The mixed structure of the SDC memristor along its switching mechanisms. (b) The U-I (voltage-current) hysteresis curve of the purchased SDC memristor. (c) The electrical characteristics test conducted on the purchased memristor. (d) The block diagram used to simulate the switching behaviour of the memristor. (e) Comparison of the actual SDC memristor and its simulation model under the same input voltage.

To design a performance-adjustable nociceptor, we first model the SDC memristor using the VTEAM model.[32] During modelling, we consider the exponential dependence of the memristance $R_M$ and the state variable $w$:

$$R_M(w) = R_{on}(\frac{R_{off}}{R_{on}})^{\frac{w-w_{on}}{w_{off}-w_{on}}} \tag{1}$$

where $R_{on}$, $R_{off}$ and $w_{on}$, $w_{off}$ are the limit values of the memristance and the state variable $w$ in the ON and OFF state, i.e., the low resistance and high resistance state. Under voltage stimuli, the state variable changes as follows:

$$\frac{dw}{dt} = \begin{cases} k_{on}(\frac{v}{v_{on}}-1)^{\alpha_{on}} f_{on}(w), & 0 < v_{on} < v \\ 0, & v_{off} < v < v_{on} \\ k_{off}(\frac{v}{v_{off}}-1)^{\alpha_{off}} f_{off}(w), & v < v_{off} < 0 \end{cases} \tag{2}$$

where $v$ is the applied voltage stimuli, $v_{on}$ and $v_{off}$ the threshold voltages, $k_{on}$, $k_{off}$, $\alpha_{on}$ and $\alpha_{off}$ constants related to the resistance switching rate of the memristor, and $f_{on}(w)$ and $f_{off}(w)$ are the windows functions to preserve the memristor state variable $w$ within the physically realistic limits. Using the quasi-newton optimization method, we optimize the parameters, including $k_{on}$ and $k_{off}$. The model parameters are listed in TABLE I. Through these steps, we establish a behaviour SDC memristor model for subsequent SPICE simulation, the corresponding block diagram and electrical characteristics are shown in FIG 2d and FIG 2e respectively.

TABLE I. The memristor parameters.

| Parameter | $R_{on}$ | $R_{off}$ | $α_{on}$ | $α_{off}$ | $k_{on}$ | $k_{off}$ | $w_{on}$ | $w_{off}$ | $v_{on}$ | $v_{off}$ |
|---|---|---|---|---|---|---|---|---|---|---|
| Value | 19 kΩ | 190 kΩ | 1 | 1 | 1.19 | -0.85 | 1 nm | 0 | 0.12 V | -0.08 V |

**2.2 Memristor-based Nociceptor System**

The designed memristor-based nociceptor system, as depicted in FIG 3a, consists of three modules: the sensory module, the signal generation module, and the memristor-based nociceptor module. Among them, the sensory module is responsible for detecting two physical stimulus - force and temperature - using piezoresistive and thermoelectric devices, respectively. The signal generation module plays a crucial role in modulating the state of the memristor based on the inputs from both the sensory module and the current status of the memristor. The finally presented memristor state indicates 'no-adaptation,' 'sensitization,' and 'recovery.' A complete processing procedure is described below. When force is applied to the piezoresistive device, the resistance change information is utilized to select pre-designed modulation schemes for the memristor, including recovery, amplification, and sensitization. The relationship between current stimuli and modulation schemes is detailed in the memristor-based nociceptor module section. In order to obtain the sensitization functionality, the current status of the memristor is also involved in determining the future modulation scheme through feedback circuitry. Therefore, both the current stimuli and the status of the memristor are key factors in modulation scheme selection.

It is worth emphasizing that the above procedure assumes a constant temperature. However, as highlighted before, temperature is a critical factor in adjusting the nociceptor performance in biological systems. In this memristor-based nociceptor system, the stimuli threshold, sensitization threshold, and recovery voltage will be adjusted based on current temperature conditions, aligning with the biological behaviours. The specific circuit design is detailed in the following sections.

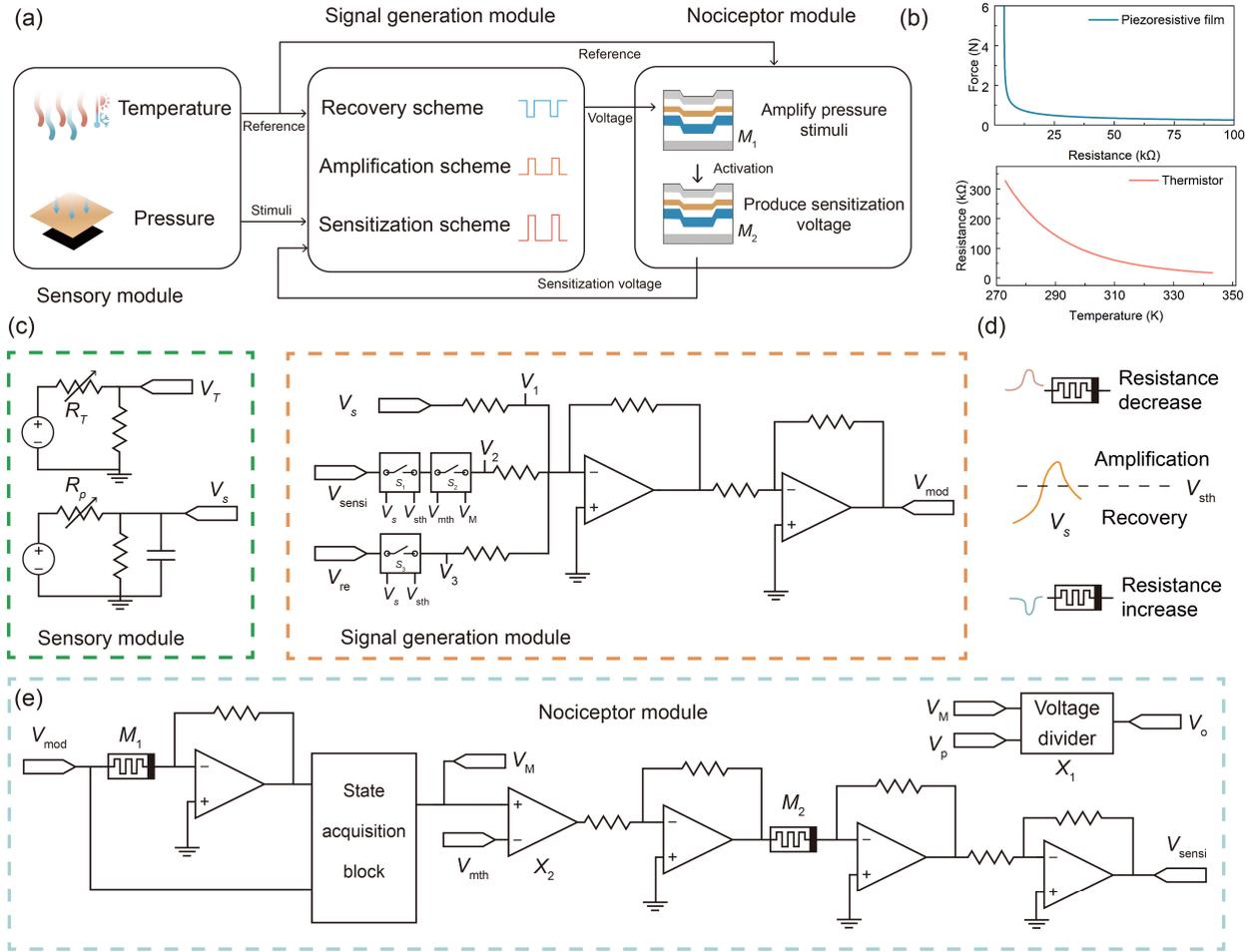

FIG.3. The memristor-based adjustable nociceptor. (a) The schematic diagram of the memristor-based adjustable nociceptor, compromising three modules: the sensory module, the signal generation module, and the memristor-based nociceptor module. (b) Response characteristics of the piezoresistive film and thermistor, which detects pressure stimuli and temperature conditions, respectively. (c) Detailed circuit design of the sensory module and signal generation module. (d) The modulation schemes. (e) Detailed circuit design of the memristor-based nociceptor module.

**Sensory Module**

The sensory module, as shown in FIG 3c, perceives external temperature information and pressure stimuli applied to the piezoresistive film. As the piezoresistive film is in series with a constant voltage source and series resistance, pressure stimuli lead to a voltage variation across the capacitance, denoted as stimuli voltage ($V_s$). Likewise, the thermistor conveys environmental temperature information through voltage variations ($V_T$) across the series resistance. FIG 3b demonstrates the response characteristics of both the pressure sensor and the temperature sensor.

**Signal Generation Module**

In the signal generation module, as shown in FIG 3c, the modulation voltage ($V_{mod}$) is produced by summing three port voltages, $V_1$, $V_2$, and $V_3$, using an op-amp-based adder circuit. In this configuration, $V_1$ is set to the modulation voltage ($V_s$). The states of analog switch $S_1$ and $S_2$ determine $V_2$. When the positive voltage ($V_M$), which is positively related to memristor resistance $M_1$ in the synaptic module, falls below the sensitization threshold ($V_{mth}$) and $V_s$ exceeds the stimuli threshold ($V_{sth}$), the sensitization modulation scheme is selected. $V_2$ is set to sensitization voltage ($V_{sensi}$), and $V_{sth}$ is reduced in this sensitization state, mimicking the sensitization feature of biological nociceptors. Otherwise, $V_2$ is set to 0 V. Similarly, the recovery voltage ($V_{re}$) is transmitted by analog switch $S_3$ only when the current stimuli voltage $V_s$ is below the threshold $V_{sth}$. The recovery modulation scheme is selected in this case, and in all other cases, $V_3$ remains 0 V. When $V_2$ and $V_3$ are set to 0 V, the amplification modulation scheme is selected as shown in FIG 3d. The memristor is modulated only by $V_s$. Consequently, the output modulation voltage $V_{mod}$ can be formulated as follows:

$$V_{mod} = V_s + V_2 + V_3 \tag{3}$$

$$V_2 = \begin{cases} V_{sensi} & V_M < V_{mth}, V_s > V_{sth} \\ 0V & otherwise \end{cases} \tag{4}$$

$$V_3 = \begin{cases} V_{re} & V_s < V_{sth} \\ 0V & otherwise \end{cases} \tag{5}$$

It is crucial to emphasize that the sensitization threshold, recovery voltage, and threshold voltage values adapt to current temperature conditions utilizing analog switches and voltage sources. The specific values at different temperatures are listed in TABLE II. Consequently, the modulation schemes will be adjusted, leading to either enhanced or depressed nociceptor responses.

TABLE II. Adjustment of threshold parameters at different temperatures.

| Parameter | 283 K | 289 K | 295 K | 301 K | 307 K |
|---|---|---|---|---|---|
| $V_{sth}$ (sensitization state) | 0.46 V | 0.33 V | 0.20 V | 0.17 V | 0.14 V |
| $V_{sth}$ (other states) | 0.56 V | 0.43 V | 0.30 V | 0.27 V | 0.24 V |
| $V_{re}$ | -0.51 V | -0.38 V | -0.25 V | -0.22 V | -0.19 V |
| $V_{mth}$ | 0.6 V | 1.3 V | 2.0 V | 2.7 V | 3.4 V |

**Memristor-Based Nociceptor Module**

As shown in FIG 3e, the memristor-based nociceptor module incorporates two memristors ($M_1$ and $M_2$), which play crucial roles in amplifying pressure stimuli and generating sensitization voltage, respectively. Initially, both memristors are in a high-resistance state. Upon selecting the modulation scheme, $M_1$ is modulated under the modulation voltage ($V_{mod}$) via a feedback op-amp circuit. The $M_1$ resistance information is then translated as a voltage $V_M$ using a memristor-state acquisition block, which mainly consists of a voltage divider circuit.[33,34] When exposed to dangerous stimuli, the positive modulation voltage leads to an increase in memristor conductance. This increasing conductance mirrors the enhanced sensitivity observed in biological nociceptors. Leveraging this similarity, the voltage divider block $X_1$ calculates the ratio voltage of the voltage ($V_p$) converted from external force to $V_M$. This output voltage not only positively correlates with the current pressure intensity but also amplifies in response to hazardous stimuli. Therefore, the resulting response voltage ($V_o$) from this block is the output of this memristor-based nociceptor, equivalent to the firing rates seen in biological nociceptors.[15,26,35] When $V_M$ falls below the threshold $V_{mth}$, the positive output voltage of the comparison circuit $X_2$ modulates $M_2$ to the low-resistance state, resulting in a sensitization voltage ($V_{sensi}$), which is sent to the signal generation module to achieve sensitization function. After removing the dangerous stimuli, $M_1$ is modulated back to a high-resistance state under the recovery voltage. Concurrently, the output voltage of $X_2$ becomes positive, reverting $M_2$ to its high-resistance state after reversing the polarity of output voltage.

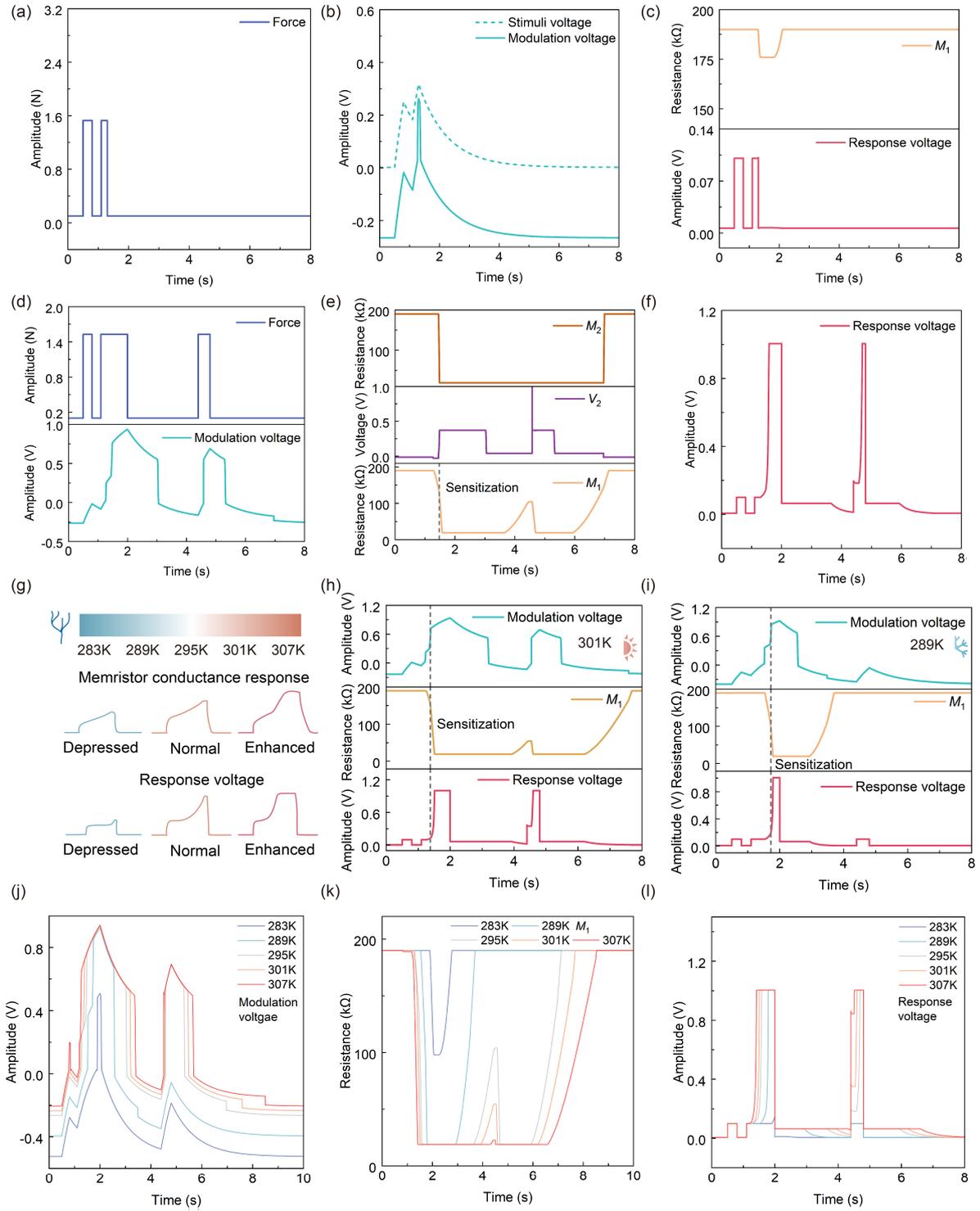

FIG.4. Functional demonstration of the adjustable memristor-based nociceptor. (a) Force stimuli applied to demonstrate the 'threshold' and 'no-adaptation' characteristics. (b) Comparison of stimuli voltage and modulation voltage under the force stimuli shown in (a). (c) Memristor response and corresponding response voltage. Upon the second force stimulus, the decrease in $M_1$ resistance is indicative of the 'no-adaptation' feature. (d) Force stimuli applied to exhibit the 'sensitization' characteristics and adaptive response at varying temperatures. (e) Memristor response and voltage at port 2. When $M_1$ falls below the threshold, $M_2$ is set to a low-resistance state, sending sensitization voltage to port 2. The increased voltage makes $M_1$ drop dramatically, observable in the changing resistance rate. (f) Response voltage of the nociceptor to the force stimuli in (d). (g) Schematic illustrations of the adaptive response of the memristor-based nociceptor at varying temperatures, showing increased sensitivity to external pressure stimuli at higher temperatures. (h) Modulation voltage, memristor response, and response voltage under the same force stimuli as in (d) at 301 K. (i) Modulation voltage, memristor response, and response voltage under the same force

stimuli as in (d) at 289 K. (j) Comparison of modulation voltages at different temperatures. (k) Comparison of memristor response at different temperatures. (l) Comparison of response voltages at different temperatures.

**2.3 Implementation of Nociceptor Functionality**

Based on the above system, we demonstrate a self-reconfigurable nociceptor with critical features, including 'threshold,' 'no-adaptation,' and 'relaxation.' The activation dynamics of biological tactile nociceptors are closely linked to the intensity, duration, and frequency of stimuli and are also sensitive to temperature variations. In our memristor-based nociceptor, the force stimuli of various amplitudes and durations are applied to demonstrate these critical features under different temperature conditions.

At a temperature of 295 K, when applying the first force stimulus shown in FIG 4a, the modulation scheme remains in the recovery state due to the weak strength of the stimulus, as depicted in FIG 4b. The $M_1$ remains in a high-resistance state. This occurs because the cumulative voltage on the capacitor does not surpass the threshold $V_{sth}$, and the recovery voltage dominates the modulation, maintaining the high resistance. Upon the application of the second force stimulus, as shown in FIG 4c, the stimuli voltage $V_s$ exceeds the threshold $V_{sth}$, and the positive modulation voltage makes the memristor resistance decrease, demonstrating the 'threshold' and 'no-adaptation' features. When external stimuli strength increases, as shown in FIG 4d, upon the second force stimulus, the modulation voltage increases dramatically at 1.48s; this is because the state of $M_1$ is set to a low-resistance state, leading to a $V_M$ below $V_{mth}$. In this case, the $M_2$ is activated, contributing an additional $V_{sensi}$ to the modulation scheme. Thus, $M_1$ in this 'injured' nociceptor drops significantly, leading to a heightened response compared to the nociceptor under the normal state, as shown in FIG 4e. At this time, the amplification of the dangerous external stimuli reaches 1000% according to the memristor change ratio (FIG 4f). This response mimics the 'sensitization' feature commonly seen in biological receptors, i.e., a nociceptor enhances the pain sensitivity when the stimulus is tensive and results in tissue damage. After removing the second force stimulus, $M_1$ is reset to the original high-resistance state under recovery voltage. Upon receiving the third force stimulus, the $M_1$ is modulated to a low-resistance state quickly, demonstrating the ability to detect potential danger.

When the environment changes, our nociceptor exhibits adaptive characteristics akin to those of biological nociceptors. As shown in FIG 4g, the response of the nociceptor is enhanced with rising temperatures and depressed when temperatures decrease. This adaptive characteristic is mainly achieved by modifying the modulation schemes, leading to $M_1$ enhanced or depressed conductance response. Thus, the response voltage alters accordingly. As depicted in FIG 4h, under the same force stimuli at 295 K, the $M_1$ activates more readily at 301 K, signifying a reduced threshold $V_{sth}$. Moreover, $V_{mth}$ is higher, loosening the criteria for entering the sensitization state. Thus, the nociceptor system enters earlier than the system at 295 K. Besides, upon stimulus removal, the recovery voltage increases, leading to an extended recovery period. At 289 K, the above parameters evolve in the opposite direction, as shown in FIG 4i, representing the depressed pain sensitivity. Notably, the third force stimulus does not activate the memristor-based nociceptor as the stimuli threshold increases. The experimental results under various temperatures are presented in FIG 4j to 4l. Drawing on the temperature information sensed by the thermal resistor, the nociceptor autonomously adjusts its response characteristics: a higher temperature induces a larger modulation voltage, an enhanced memristor conductance response, and an enhanced response voltage. The memristor response varies significantly, exhibiting a difference ratio of up to 500% between 283 K and 307 K. Compared to current state-of-the-art neuromorphic nociceptors, our results highlight the exploitation of diverse sensory modalities, achieving self-adjusting characteristics that adapt seamlessly to varying environmental conditions, as shown in TABLE III.

TABLE III. Comparison with other works.

|  | This work | Yoon et al.[11] | John et al.[25] | Liu et al.[27] | Xu et al.[35] | Im et al.[36] |
|---|---|---|---|---|---|---|
| Sensory modalities | 2 types | 1 type | 1 type | 1 type | 1 type | 1 type |
| Applicable device conditions | Nonlinear | Nonlinear | Nonlinear | Nonlinear | Nonlinear | Nonlinear |
|  |  | Volatile | Volatile | Volatile | Volatile | Volatile |
| Threshold function | √ | √ | √ | √ | √ | √ |
| No-adaptation function | √ | √ | √ | √ | √ | √ |
| Sensitization function | √ | √ | √ | × | √ | √ |
| Recovery function | √ | √ | √ | √ | √ | √ |
| Self-adjustable function | √ | × | × | × | × | × |

## 3. Conclusion

Perception threshold self-adjustment is an essential functionality of biological nociceptors for human beings to adapt well to external dynamic environments. This is now also a must-have functionality for intelligent robotics to both implement dangerous tasks and protect mechanical structures in unstructured conditions. Compared to traditional technologies that utilize complex algorithms to examine the status of robotics and make operational decisions, this article provides an example of how to achieve the same goal, albeit in a much simpler manner. By employing the memristor's intrinsic similarity to biological receptors and the sensors read pressure and temperature information as the modulation scheme indicators, a self-reconfigurable artificial nociceptor is assembled to automatically change its harm threshold under different environmental conditions. Notably, this design is universally applicable across memristors, independent of their specific switching mechanisms, presenting a versatile solution for wide neuromorphic computing applications. This universality is evidenced by the use of a commercial

SDC memristor in the design. Moreover, the principles outlined in this article can be further extended into various senses, leading to a comprehensive perception system, hence strongly supporting the development of intelligent robotics.

## 4. Experimental Section

*Electrical Characterizations*: To measure the electrical characteristics of the SDC memristor, we employed a Tektronix Keithley 4200-SCS parameter analyzer equipped with pulse measurement units. This setup was utilized to obtain the memristor's hysteresis curve and pulse response

**Conflicts of interest**

There are no conflicts to declare.

**Acknowledgements**

S.G. acknowledges support from the National Key Research and Development Program of China (grant No. 2023YFB3208003), National Natural Science Foundation of China (grant No. 62171014), and Beihang University (grants No.KG161250 and ZG16S2103). G.H.H. acknowledges support from RGC (24200521) and SHIAE (RNEp3-21).